\pdfoutput=1
\documentclass[sigconf]{acmart}

\AtBeginDocument{%
  \providecommand\BibTeX{{%
    \normalfont B\kern-0.5em{\scshape i\kern-0.25em b}\kern-0.8em\TeX}}}

\copyrightyear{2021}
\acmYear{2021}
\setcopyright{rightsretained}
\acmConference[AIES '21]{Proceedings of the 2021 AAAI/ACM Conference on AI, Ethics, and Society}{May 19--21, 2021}{Virtual Event, USA}
\acmBooktitle{Proceedings of the 2021 AAAI/ACM Conference on AI, Ethics, and Society (AIES '21), May 19--21, 2021, Virtual Event, USA}
\acmISBN{978-1-4503-8473-5/21/05}
\acmDOI{10.1145/3461702.3462540}

\begin{document}
\fancyhead{}

\title{Fairness for Unobserved Characteristics:\\Insights from Technological Impacts on Queer Communities}

\author{Nenad Tomasev}
\email{nenadt@deepmind.com}
\orcid{0000-0003-1624-0220}
\affiliation{%
  \institution{DeepMind}
  \city{London}
  \country{United Kingdom}
}

\author{Kevin R. McKee}
\email{kevinrmckee@deepmind.com}
\orcid{0000-0002-4412-1686}
\affiliation{%
  \institution{DeepMind}
  \city{London}
  \country{United Kingdom}
}

\author{Jackie Kay}
\email{kayj@deepmind.com}
\affiliation{%
  \institution{DeepMind}
  \city{London}
  \country{United Kingdom}
}
\additionalaffiliation{%
  \institution{Centre for Artificial Intelligence, University College London}
  \city{London}
  \country{United Kingdom}
}

\author{Shakir Mohamed}
\email{shakir@deepmind.com}
\orcid{0000-0002-1184-5776}
\affiliation{%
  \institution{DeepMind}
  \city{London}
  \country{United Kingdom}
}

\begin{abstract}
Advances in algorithmic fairness have largely omitted sexual orientation and gender identity. We explore queer concerns in privacy, censorship, language, online safety, health and employment to study the positive and negative effects of artificial intelligence on queer communities. These issues underscore the need for new directions in fairness research that take into account a multiplicity of considerations, from privacy preservation, context sensitivity and process fairness, to an awareness of sociotechnical impact and the increasingly important role of inclusive and participatory research processes.  
Most current approaches for algorithmic fairness assume that the target characteristics for fairness---frequently, race and legal gender---can be observed or recorded. Sexual orientation and gender identity are prototypical instances of \textit{unobserved} characteristics, which are frequently missing, unknown or fundamentally unmeasurable. This paper highlights the importance of developing new approaches for algorithmic fairness that break away from the prevailing assumption of observed characteristics.
\end{abstract}



\maketitle

\section{Introduction}

As the field of algorithmic fairness has matured, the ways in which machine learning researchers and developers operationalise approaches for fairness have expanded in scope and applicability. 
Fairness researchers have made important advances and demonstrated how the risks of algorithmic systems are imbalanced across different characteristics of the people who are analysed and affected by classifiers and decision-making systems~\citep{barocas2019fairness, fudenberg2012fairness}. Progress has been particularly strong with respect to race and legal gender.\footnote{Throughout this paper, we distinguish between `legal gender' (the gender recorded on an individual's legal documents, often assigned to them at birth by the government, physicians or their parents) and `gender identity' (an individual's personal feelings and convictions about their gender; \citep{brookgender}).} Fairness studies have helped to draw attention to racial bias in recidivism prediction~\citep{angwin2016machine}, expose racial and gender bias in facial recognition \citep{buolamwini2018gender}, reduce gender bias in language processing~\citep{bolukbasi2016man, park2018reducing}, and increase the accuracy and equity of decision making for child protective services~\citep{chouldechova2018case}.

Algorithms have moral consequences for queer communities, too.
However, algorithmic fairness for queer individuals and communities remains critically underexplored.
In part, this stems from the unique challenges posed by studying sexual orientation and gender identity.
Most definitions of algorithmic fairness share a basis in norms of egalitarianism~\citep{barocas2019fairness, binns2018fairness}.\footnote{It is worth noting that certain algorithmic domains supplement egalitarian concerns with additional ethical values and principles. For example, fairness assessments of healthcare applications typically incorporate beneficence and non-malfeasance, two principles central to medical ethics~\citep{beauchamp2001principles}.} For example, classification parity approaches to fairness aim to equalise predictive performance measures across groups, whereas anti-classification parity approaches rely on the omission of protected attributes from the decision making process to ensure different groups receive equivalent treatment~\citep{corbett2018measure}. An inherent assumption of these approaches is that the protected characteristics are known and available within datasets. Sexual orientation and gender identity are prototypical examples of \textit{unobserved} characteristics, presenting challenging obstacles for fairness research~\citep{andrus2020we, jacobs2019measurement}.

This paper explores the need for \textit{queer fairness} by reviewing the experiences of technological impacts on queer communities. For our discussion, we define `queer' as `possessing non-normative sexual identity, gender identity, and/or sexual characteristics'. We consider this to include lesbian, gay, bisexual, pansexual, transgender, and asexual identities---among others.\footnote{Throughout this paper, we use `queer' and `LGBTQ+' interchangeably. The heterogeneity of queer communities---and the complexity of the issues they face---preclude this work from being an exhaustive review of queer identity. As a result, there are likely perspectives that were not included in this manuscript, but that have an important place in broader discussions of queer fairness.}

The focus on queer communities is important for several reasons. 
Given the historical oppression and contemporary challenges faced by queer communities, there is a substantial risk that artificial intelligence (AI) systems will be designed and deployed unfairly for queer individuals.
Compounding this risk, sensitive information for queer people is usually not available to those developing AI systems, rendering the resulting unfairness unmeasurable from the perspective of standard group fairness metrics. 
Despite these issues, fairness research with respect to queer communities is an understudied area. Ultimately, the experiences of queer communities can reveal insights for algorithmic fairness that are transferable to a broader range of characteristics, including disability, class, religion, and race.

This paper aims to connect ongoing efforts to strengthen queer communities in AI research~\citep{queerinaineurips2018, queerinaineurips2019, queerinaiicml2020} and sociotechnical decision making~\citep{outintech, lesbianswhotech, intertech, lgbtinstitute} with recent advances in fairness research, including promising approaches to protecting unobserved characteristics. This work additionally advocates for the expanded inclusion of queer voices in fairness and ethics research, as well as the broader development of AI systems. We make three contributions in this paper:
\begin{enumerate}
    \item Expand on the promise of AI in empowering queer communities and supporting LGBTQ+ rights and freedoms. 
    \item Emphasise the potential harms and unique challenges raised by the sensitive and unmeasurable aspects of identity data for queer people.
    \item Based on use cases from the queer experience, establish requirements for algorithmic fairness on unobserved characteristics.
\end{enumerate}

\section{Considerations for Queer Fairness}

To emphasise the need for in-depth study of the impact of AI on queer communities around the world, we explore several interconnected case studies of how AI systems interact with sexual orientation and gender identity. These case studies highlight both potential benefits and risks of AI applications for queer communities. In reviewing these cases, we hope to motivate the development of technological solutions that are inclusive and beneficial to everyone. Importantly, these case studies will demonstrate cross-cutting challenges and concerns raised by unobserved and missing characteristics, such as preserving privacy, supporting feature imputation, context-sensitivity, exposing coded inequity, participatory engagement, and sequential and fair processes. %

\subsection{Privacy}

Sexual orientation and gender identity are highly private aspects of personal identity. Outing queer individuals---by sharing or exposing their sexual orientation or gender identity without their prior consent---can not only lead to emotional distress, but also risk serious physical and social harms, especially in regions where queerness is openly discriminated against~\citep{wang2019discrimination}, criminalised~\citep{dejong2014death} or persecuted~\citep{scicchitano2019real}. Privacy violations can thus have major consequences for queer individuals, including infringement upon their basic human rights~\citep{bosia2020oxford, amnestyinternational2015lgbti}, denial of employment and education opportunities, ill-treatment, torture, sexual assault, rape, and extrajudicial killings.

\subsubsection{Promise}

Advances in privacy-preserving machine learning~\citep{nasr2018machine, bonawitz2017practical, jayaraman2019evaluating} present the possibility that the queer community might benefit from AI systems while minimising the risk of information leakage. Researchers have proposed adversarial filters~\citep{zhang2020adversarial, liu2017protecting} to obfuscate sensitive information in images and speech shared online while reducing the risks of re-identification.

Still, challenges remain~\citep{srivastava2019privacy}. More research is needed to ensure the robustness of the adversarial approaches. The knowledge gap between the privacy and machine-learning research communities must be bridged for these approaches to achieve the desired effects. This will ensure that the appropriate types of protections are included in the ongoing development of AI solutions~\citep{al2019privacy}.

\subsubsection{Risks}

A multitude of privacy risks arise for queer people from the applications of AI systems. We focus in particular on the categorisation of identity from sensitive data, the ethical risk of surveillance, and invasions of queer spaces.

In 2017, Stanford researchers attempted to build an AI `gaydar', a computer vision model capable of guessing a person's sexual orientation from images~\citep{wang2018deep}. The resulting algorithm, a logistic regression model trained on 35,326 facial images, achieved a high reported accuracy in identifying self-reported sexual orientation across both sexes. The results of this study have since been questioned, largely on the basis of a number of methodological and conceptual flaws that discredit the performance of the system~\citep{gelman2018gaydar}.
Other algorithms designed to predict sexual orientation have suffered similar methodological and conceptual deficiencies.
A recently released app claimed to be able to quantify the evidence of one's non-heterosexual orientation based on genetic data, for example~\citep{bellenson2019122}, largely obfuscating the limited ability of genetic information to predict sexual orientation~\citep{ganna2019large}.

Though these specific efforts have been flawed, it is plausible that in the near future algorithms could achieve high accuracy, depending on the data sources involved.
Behavioural data recorded online present particular risks to the privacy of sexual orientation and gender identity: after all, the more time people spend online, the greater their digital footprint. AI `gaydars' relying on an individual's recorded interests and interactions could pose a serious danger to the privacy of queer people.
In fact, as a result of the long-running perception of the queer community as a profitable `consumer group' by business and advertisers alike, prior efforts have used online data to map `queer interests' in order to boost sales and increase profits~\citep{sender2018gay}. In at least one instance, researchers have attempted to use basic social media information to reconstruct the sexual orientation of users~\citep{bhattasali2015machine}.

The ethical implications of developing such systems for queer communities are far-reaching, with the potential of causing serious harms to affected individuals. Prediction algorithms could be deployed at scale by malicious actors, particularly in nations where homosexuality and gender nonconformity are punishable offences. %
In fact, in many such nations, authorities already use technology to entrap or locate queer individuals through social media and LGBTQ+ dating apps (e.g., \citep{culzac2014egypt}). Systems predicting sexual orientation may also exacerbate the pre-existing privacy risks of participating in queer digital spaces. There have been recorded cases of coordinated campaigns for outing queer people, resulting in lives being ruined, or lost due to suicide~\citep{embury2020bullied}. These malicious outing campaigns have until now been executed at smaller scales. However, recent developments in AI greatly amplify the potential scale of such incidents, endangering larger communities of queer people in certain parts of the world. Facial recognition technology~\citep{voulodimos2018deep} could be employed by malicious actors to rapidly identify individuals sharing their pictures online, whether publicly or in direct messages. Facial recognition could similarly be used to automatically identify people in captured recordings of protests, in queer nightclubs or community spaces, and other in-person social events. These possibilities highlight the potential dangers of AI for state-deployed surveillance technology. Chatbots have similarly been deployed to elicit private information on dating apps, compromising users' device integrity and privacy~\citep{mccormick2015warning}. Existing bots are scripted, and therefore can usually be distinguished from human users after longer exchanges. Nonetheless, strong language models~\citep{brown2020language} threaten to exacerbate such privacy risks, given their ability to quickly adjust to the style of communication based on a limited number of examples. These language models amplify existing concerns around the collection of private information and the compromising of safe online spaces.

In addition to the direct risks to privacy, algorithms intended to predict sexual orientation and gender identity also perpetuate concerning ideas and beliefs about queerness.
Systems using genetic information as the primary input, for example, threaten to reinforce biological essentialist views of sexual orientation and echo tenets of eugenics---a historical framework that leveraged science and technology to justify individual and structural violence against people perceived as inferior~\citep{ordover2003american, wolbring2001we}. More broadly, the design of predictive algorithms can lead to erroneous beliefs that biology, appearance or behaviour are the essential features of sexual orientation and gender identity, rather than imperfectly correlated causes, effects or covariates of queerness.

In sum, sexual orientation and gender identity are associated with key privacy concerns. Non-consensual outing and attempts to infer protected characteristics from other data thus pose ethical issues and risks to physical safety. In order to ensure queer algorithmic fairness, it will be important to develop methods that can improve fairness for marginalised groups without having direct access to group membership information. This could be achieved either through proxies~\citep{gupta2018proxy} when there is sufficient contextual information in the data, or by implementing more general principles to ensure that similar individuals are treated similarly~\citep{dwork2012fairness}, under any plausible unobserved grouping~\citep{lahoti2020fairness}.

\subsection{Censorship}

Multiple groups and institutions around the world impose unjust restrictions on the freedom of expression and speech of queer communities. This censorship is often justified by its supporters as `preserving decency' and `protecting the youth', but in reality leads to the erasure of queer identity. Laws against `materials promoting homosexuality' were established in the late 1980s in the United Kingdom and repealed as late as 2003~\citep{burridge2004not}. Nations that are considered major world powers have laws banning the portrayal of same-sex romances in television shows (e.g., China; \citep{lu2016china}), the mention of homosexuality or transgender identities in public education (e.g., state-level laws in the United States; \citep{hoshall2012afraid}), or any distribution of LGBT-related material to minors (e.g., Russia; \citep{kondakov2019censorship}). Not only do such laws isolate queer people from their communities---particularly queer youth---they shame queerness as indecent behaviour, setting a precedent for further marginalisation and undermining of human rights. 
Many queer content producers in such nations have argued that their online content is being restricted and removed at the detriment of queer expression and sex positivity, as well as at the cost of their income~\citep{york2015privatising}.

\subsubsection{Promise}

AI systems may be effectively used to mitigate censorship of queer content. Machine learning has been used to analyse and reverse-engineer patterns of censorship. A study of 20 million tweets from Turkey employed machine learning to show that the vast majority of censored tweets contained political content~\citep{tanash2015known}. A statistical analysis of Weibo posts and Chinese-language tweets uncovered a set of charged political keywords present in posts with anomalously high deletion rates~\citep{bamman2012censorship}. Further study of censorship could be key to drawing the international community's attention to human rights violations. It might also be useful for empowering affected individuals to circumvent these unfair restrictions. For example, a comparative analysis of censorship across platforms might help marginalised communities identify safe spaces where freedom of expression is least obstructed, as well as provide evidence and help coordinate action against platforms responsible for discriminatory censorship.
Nonetheless, no large-scale study of censored queer content has yet been conducted, rendering these forward-looking technical applications somewhat speculative at the moment. %

\subsubsection{Risk}

Although we believe machine learning can be used to combat censorship, tools for detecting queer digital content can be abused to enforce censorship laws or heteronormative cultural attitudes. As social network sites, search engines, and other media platforms adopt algorithms to moderate content at scale, the risk for unfair or biased censorship of queer content increases, and governing entities are empowered to erase queer identities from the digital sphere~\citep{cobbe2019algorithmic}. Automated content moderation systems are at risk of censoring queer expression even when the intention is benign, such as protecting users from verbal abuse. %
To help combat censorship restrictions and design fair content moderation systems, ML fairness researchers could investigate how to detect and analyse anomalous omission of information related to queer identity (or other protected characteristics) in natural language and video data.

Censorship often goes hand-in-hand with the distortion of facts. Recent advances in generative models have made the fabrication of digital content trivial, given enough data and computational power~\citep{chesney2019deep}. 
Malicious and dehumanising misinformation about the queer community has been used as justification for abuse and suppression throughout history, tracing back to medieval interpretations of ancient religious texts~\citep{dynes2014homophobic}. Technological and political solutions to the threat of misinformation are important for protecting queer expression---as well as global democracy. The AI community has begun to develop methods to verify authentic data through, for example, open datasets and benchmarks for detecting synthetic images and video~\citep{rossler2019faceforensics}.

While the goal of fairness for privacy is preventing the imputation of sensitive data, the goal of fairness for censorship is to reveal the unfair prevention of expression. This duality could surface important technical connections between these fields. In terms of social impact, many people around the world outside of the queer community are negatively affected by censorship. Further research in fairness for censorship could have far-reaching benefit across technical fields, social groups and borders.

\subsection{Language}

Language encodes and represents our way of thinking and communicating about the world. There is a long history of oppressive language being weaponised against the queer community~\citep{nadal2011sexual, thurlow2001naming}, highlighting the need for developing fair and inclusive language models.

Inclusive language~\citep{weinberg2009lgbt} extends beyond the mere avoidance of derogatory terms, as there are many ways in which harmful stereotypes can surface. For example, the phrase `That's so gay'~\citep{chonody2012that} equates queerness with badness. Using the term `sexual preference' rather than `sexual orientation' can imply that sexual orientation is a volitional choice, rather than an intrinsic part of one's identity.
Assuming one's gender identity, without asking, is harmful to the trans community as it risks misgendering people. This can manifest in the careless use of assumed pronouns, without knowledge of an individual's identification and requested pronouns. Reinforcing binary and traditional gender expression stereotypes, regardless of intent, can have adverse consequences. %
The use of gender-neutral pronouns has been shown to result in lower bias against women and LGBTQ+ people~\citep{tavits2019language}. To further complicate the matter, words which originated in a derogatory context, such as the label `queer' itself, are often reclaimed by the community in an act of resistance. This historical precedent suggests that AI systems must be able to adapt to the evolution of natural language and avoid censoring language based solely on its adjacency to the queer community.

\subsubsection{Promise}

Natural language processing applications permeate the field of AI. These applications include use cases of general interest like machine translation, speech recognition, sentiment analysis, question answering, chatbots and hate speech detection systems. There is an opportunity to develop language-based AI systems inclusively---to overcome human biases and establish inclusive norms that would facilitate respectful communication with regards to sexual orientation and gender identity~\citep{strengers2020adhering}.

\subsubsection{Risks}

Biases, stereotypes and abusive speech are persistently present in top-performing language models, as a result of their presence in the vast quantities of training data that are needed for model development~\citep{costa2019analysis}. Formal frameworks for measuring and ensuring fairness~\citep{hendrycks2020aligning, hendrycks2020measuring, sheng2020towards} in language are still in nascent stages of development. Thus, for AI systems to avoid reinforcing harmful stereotypes and perpetuating harm to marginalised groups, research on inclusive language requires more attention. For language systems to be fair, they must be capable of reflecting the contextual nature of human discourse.

\subsection{Fighting Online Abuse}
\label{subsection:onlineabuse}

The ability to safely participate in online platforms is critical for marginalised groups to form a community and find support~\citep{yiling2020how}. However, this is often challenging due to pervasive online abuse~\citep{jane2020online}.
Queer people are frequently targets of internet hate speech, harassment and trolling. This abuse may be directed at the community as a whole or at specific individuals who express their queer identity online. Adolescents are particularly vulnerable to cyberbullying and the associated adverse effects, including depression and suicidal ideation~\citep{abreu2018cyberbullying}. Automated systems for moderation of online abuse are a possible solution that can protect the psychological safety of the queer community at a global scale.

\subsubsection{Promise}

AI systems could potentially be used to help human moderators flag abusive online content and communication directed at members of marginalised groups, including the queer community~\citep{saha2019hatemonitors, schmidt2017survey}. A proof of concept for this application was developed in the Troll Patrol project~\citep{delisle2019large, amnestyinternationaltroll}, a collaboration between Amnesty International and Element AI's former AI for Good team. The Troll Patrol project investigated the application of natural language processing methods for quantifying abuse against women on Twitter. The project revealed concerning patterns of online abuse and highlighted the technological challenges required to develop online abuse detection systems. Recently, similar systems have been applied to tweets directed at the LGBTQ+ community. Machine learning and sentiment analysis were leveraged to predict homophobia in Portuguese tweets, resulting in 89.4\% accuracy~\citep{pereira2018using}. Deep learning has also been used to evaluate the level of public support and perception of LGBTQ+ rights following the Supreme Court of India's verdict regarding the decriminalisation of homosexuality~\citep{khatua2019tweeting}.

The ways in which abusive comments are expressed when targeted at the trans community pose some idiosyncratic research challenges. In order to protect the psychological safety of trans people, it is necessary for automated online abuse detection systems to properly recognise acts of misgendering (referring to a trans person with the gender they were assigned at birth) or `deadnaming' (referring to a trans person by the name they had before they transitioned; \citep{stonewalltruth}). These systems have a simultaneous responsibility to ensure that deadnames and other sensitive information are kept private to the user. It is therefore essential for the queer community to play an active role in informing the development of such systems.

\subsubsection{Risks}

Systems developed with the purpose of automatically identifying toxic speech could introduce harms by failing to recognise the context in which speech occurs. Mock impoliteness, for example, helps queer people cope with hostility; the communication style of drag queens in particular is often tailored to be provocative. A recent study~\citep{gomes2019drag} demonstrated that an existing toxicity detection system would routinely consider drag queens to be as offensive as white supremacists in their online presence. The system further specifically associated high levels of toxicity with words like `gay', `queer' and `lesbian'.

Another risk in the context of combating online abuse is unintentionally disregarding entire groups through ignorance of intersectional issues. Queer people of colour experience disproportionate exposure to online (and offline) abuse~\citep{balsam2011measuring}, even within the queer community itself. This imbalance stems from intersectional considerations about the ways in which race, class, gender, and other individual characteristics may combine into differential modes of discrimination and privilege (``intersectionality''; \citep{crenshaw1989demarginalizing}). Neglecting intersectionality can lead to disproportionate harms for such subcommunities.

To mitigate these concerns, it is important for the research community to employ an inclusive and participatory approach~\citep{martin2020participatory} when compiling training datasets for abusive speech detection. For example, there are homophobic and transphobic slurs with a racialised connotation that should be included in training data for abuse detection systems. Furthermore, methodological improvements may help advance progress. Introducing fairness constraints to model training has demonstrably helped mitigate the bias of cyber-bullying detection systems~\citep{gencoglu2020cyberbullying}. Adversarial training can similarly assist by demoting the confounds associated with texts of marginalised groups~\citep{xia2020demoting}.

\subsection{Health}

The drive towards equitable outcomes in healthcare entails a set of unique challenges for marginalised communities. Queer communities have been disproportionately affected by HIV~\citep{singh2018hiv}, suffer a higher incidence of sexually-transmitted infections, and are afflicted by elevated rates of substance abuse~\citep{wallace2017addictions}. Compounding these issues, queer individuals frequently experience difficulties accessing appropriate care~\citep{bize2011access, humanrightswatch2018us}. %
Healthcare professionals often lack appropriate training to best respond to the needs of LGBTQ+ patients~\citep{schneider2019glma}. Even in situations where clinicians do have the proper training, patients may be reluctant to reveal their sexual orientation and gender identity, given past experiences with discrimination and stigmatisation.

In recent months, the COVID-19 pandemic has amplified health inequalities~\citep{bowleg2020we, van2020covid}. Initial studies during the pandemic have found that LGBTQ+ patients are experiencing poorer self-reported health compared to cisgendered heterosexual peers~\citep{oneill2020health}. The health burden of COVID-19 may be especially severe for queer people of colour, given the substantial barriers they face in accessing healthcare~\citep{hsieh2016sexual}.

\subsubsection{Promise}

To this day, the prevalence of HIV among the queer community remains a major challenge. Introducing systems that both reduce the transmission risk and improve care delivery for HIV+ patients will play a critical role in improving health outcomes for queer individuals.

Machine learning presents key opportunities to augment medical treatment decisions~\citep{bisaso2017survey}. For example, AI may be productively applied to identify the patients most likely to benefit from pre-exposure prophylaxis for HIV. A research team recently developed such a system, which correctly identified 38.6\% of future cases of HIV~\citep{marcus2019use}. The researchers noted substantial challenges: model sensitivity on the validation set was 46.4\% for men and 0\% for women, highlighting the importance of intersectionality for fair outcomes in healthcare.
Machine learning has also been used to predict early virological suppression~\citep{bisaso2018comparative}, adherence to anti-retroviral therapy~\citep{semerdjian2018supervised}, and individual risk of complications such as chronic kidney disease~\citep{roth2020cohort} or antiretroviral therapy-induced mitochondrial toxicity~\citep{lee2019comparison}.

\subsubsection{Risks}

Recent advances in AI in healthcare may lead to widespread increases in welfare. Yet there is a risk that benefits will be unequally distributed---and an additional risk that queer people's needs will not be properly met by the design of current systems. Information about sexual orientation and gender identity is frequently absent from research datasets. To mitigate the privacy risk for patients and prevent reidentification, HIV status and substance abuse are also routinely omitted from published data. While such practices may be necessary, it is worth recognising the important downstream consequences they have for AI system development in healthcare. It can become impossible to assess fairness and model performance across the omitted dimensions. Moreover, the unobserved data increase the likelihood of reduced predictive performance (since the features are dropped), which itself results in worse health outcomes. The coupled risk of a decrease in performance and an inability to measure it could drastically limit the benefits from AI in healthcare for the queer community, relative to cisgendered heterosexual patients. To prevent the amplification of existing inequities, there is a critical need for targeted fairness research examining the impacts of AI systems in healthcare for queer people.

To help assess the quality of care provided to LGBTQ+ patients, there have been efforts aimed at approximately identifying sexual orientation~\citep{bjarnadottir2019nurse} and gender identity~\citep{ehrenfeld2019development} from clinical notes within electronic health record systems. While well intentioned, these machine learning models offer no guarantee that they will only identify patients who have explicitly disclosed their identities to their healthcare providers. These models thus introduce the risk that patients will be outed without their consent. Similar risks arise from models developed to rapidly identify HIV-related social media data~\citep{young2017toward}.

The risk presented by AI healthcare systems could potentially intensify during medical gender affirmation. %
There are known adverse effects associated with transition treatment~\citep{moore2003endocrine}. The active involvement of medical professionals with experience in cross-sex hormonal therapy is vital for ensuring the safety of trans people undergoing hormone therapy or surgery. Since cisgendered individuals provide the majority of anonymised patient data used to develop AI systems for personalised healthcare, there will be comparatively fewer cases of trans patients experiencing many medical conditions. This scarcity could have an adverse impact on model performance---there will be an insufficient accounting for the interactions between the hormonal treatment, its adverse effects and potential comorbidities, and other health issues potentially experienced by trans patients.

Framing fairness as a purely technical problem that can be addressed by the mere inclusion of more data or computational adjustments is ethically problematic, especially in high-stakes domains like healthcare~\citep{mccradden2020ethical}. Selection bias and confounding in retrospective data make causal inference particularly hard in this domain. Counterfactual reasoning may prove key for safely planning interventions aimed at improving health outcomes~\citep{prosperi2020causal}. It is critical for fairness researchers to engage deeply with both clinicians and patients to ensure that their needs are met and AI systems in healthcare are developed and deployed safely and fairly.

\subsection{Mental Health}

Queer people are more susceptible to mental health problems than their heterosexual and cisgender peers, largely as a consequence of the chronically high levels of stress associated with prejudice, stigmatisation and discrimination~\citep{meyer1995minority, meyer2003prejudice, mays2001mental, tebbe2016suicide}.
As a result, queer communities experience substantial levels of anxiety, depression and suicidal ideation~\citep{mentalhealthfoundation2020mental}. Compounding these issues, queer people often find it more difficult to ask for help and articulate their distress~\citep{mcdermott2015asking} and face systemic barriers to treatment~\citep{romanelli2017individual}. A recent LGBTQ+ mental health survey highlighted the shocking extent of issues permeating queer communities~\citep{trevorproject2020national}: 40\% of LGBTQ+ respondents seriously considered attempting suicide in the past twelve months, with more than half of transgender and nonbinary youth having seriously considered suicide; 68\% of LGBTQ+ youth reported symptoms of generalised anxiety disorder in the past two weeks, including more than three in four transgender and nonbinary youth; 48\% of LGBTQ+ youth reported engaging in self-harm in the past twelve months, including over 60\% of transgender and nonbinary youth.

\subsubsection{Promise}
AI systems have the potential to help address the alarming prevalence of suicide in the queer community. Natural language processing could be leveraged to help human operators identify cases of an increased suicide risk more reliably, and respond to them appropriately. These predictions could be based either on traditional data sources such as questionnaires and recorded interactions with mental health support workers, or new data sources including social media and engagement data. The Trevor Project, a prominent American organisation providing crisis intervention and suicide prevention services to LGBTQ+ youth~\citep{trevorproject}, is working on such an initiative. The crisis contact simulator developed by The Trevor Project team has been designed to emulate plausible conversations and interactions between the helpline workers and the callers. The Trevor Project uses this simulator to help train new team members, allowing them to practice their skills. The narrow focus of this application mitigates some of the risks intrinsic to the application of language models, though the effectiveness of the system is still being evaluated. In partnership with Google.org and its research fellows, The Trevor Project also developed an AI system to identify and prioritise community members at high risk while simultaneously increasing outreach to new contacts. The system was designed to relate different types of intake-form responses to downstream diagnosis risk levels. A separate group of researchers developed a language processing system~\citep{liang2019clustering} to identify help-seeking conversations on LGBTQ+ support forums, with the aim of helping at-risk individuals manage and overcome their issues.

In other healthcare contexts, reinforcement learning has recently demonstrated potential in steering behavioural interventions~\citep{yom2017encouraging} and improving health outcomes. Reinforcement learning represents a natural framework for personalised health interventions, since it can be set up to maximise long-term physical and mental well-being~\citep{tabatabaei2018narrowing}. If equipped with natural language capabilities, such systems might be able to act as personalised mental health assistants empowered to support mental health and escalate situations to human experts in concerning situations.

\subsubsection{Risks}

Substantial risks accompany these applications. Overall, research on any intervention-directed systems should be undertaken in partnership with trained mental health professionals and organisations, given the considerable risks associated with misdiagnosing mental illness (cf. \citep{suite2007beyond}) and exacerbating the vulnerability of those experiencing distress.

The automation of intervention decisions and mental health diagnoses poses a marked risk for the trans community. In most countries, patients must be diagnosed with gender dysphoria---an extensive process with lengthy wait times---before receiving treatments such as hormone therapy or surgery (e.g., \citep{nationalhealthservicegender}). During this process, many transgender individuals experience mistrust and invalidation of their identities from medical professionals who withhold treatment based on rigid or discriminatory view of gender~\citep{ashley2019gatekeeping}. Automating the diagnosis of gender dysphoria may recapitulate these biases and deprive many transgender patients of access to care. %

Mental health information is private and sensitive. While AI systems have the potential to aid mental health workers in identifying at-risk individuals and those who would most likely benefit from intervention, such models may be misused in ways that expose the very people they were designed to support. Such systems could also lead queer communities to be shut out from employment opportunities or to receive higher health insurance premiums. Furthermore, reinforcement learning systems for behavioural interventions will present risks to patients unless many open problems in the field can be resolved, such as safe exploration~\citep{hans2008safe} and reward specification~\citep{krakovna2020specification}. %
The development of safe intervention systems that support the mental health of the queer community is likely also contingent on furthering frameworks for sequential fairness~\citep{heidari2018preventing}, to fully account for challenges in measuring and promoting queer ML fairness.  

\subsection{Employment}

Queer people often face discrimination both during the hiring process (resulting in reduced job opportunities) and once hired and employed (interfering with engagement, development and well-being; \citep{sears2011documented}). Non-discrimination laws and practices have had a disparate impact across different communities. Employment nondiscrimination acts in the United States have led to an average increase in the hourly wages of gay men by 2.7\% and a decrease in employment of lesbian women by 1.7\%~\citep{burn2018not}, suggesting that the impact of AI on employment should be examined through an intersectional lens.

\subsubsection{Promise}

To effectively develop AI systems for hiring, researchers must first attempt to formalise a model of the hiring process. Formalising such models may make it easier to inspect current practices and identify opportunities for removing existing biases (e.g., \citep{li2020hiring}). Incorporating AI into employment decision processes could potentially prove beneficial if unbiased systems are developed~\citep{houser2019can}, though this seems difficult at the present moment and carries serious risks.

\subsubsection{Risks}

Machine learning-based decision making systems (e.g., candidate prioritisation systems) developed using historical data could assign lower scores to queer candidates, purely based on historical biases. Prior research has demonstrated that resumes containing items associated with queerness are scored significantly lower by human graders than the same resumes with such items removed~\citep{lecroy2019influence}. These patterns can be trivially learned and reproduced by resume-parsing machine learning models.

A combination of tools aimed at social media scraping, linguistic analysis, and an analysis of interests and activities could indirectly infringe of candidates' privacy by outing them to their prospective employers without their prior consent. The interest in these tools stems from the community's emphasis on big data approaches, not all of which will have been scientifically verified from the perspective of impact on marginalised groups.

Both hiring and subsequent employment are multi-stage processes of considerable complexity, wherein technical AI tools may be used across multiple stages. Researchers will not design and develop truly fair AI systems by merely focusing on metrics of subsystems in the process, abstracting away the social context of their application and their interdependence. It is instead necessary to see these as \textit{sociotechnical systems} and evaluate them as such~\citep{selbst2019fairness}.

\section{Sources of Unobserved Characteristics}

Most algorithmic fairness studies have made progress because of their focus on \textit{observed} characteristics---commonly, race and legal gender.
To be included in training or evaluation data for an algorithm, an attribute must be measured and recorded.
Many widely available datasets thus focus on immutable characteristics (such as ethnic group) or characteristics which are recorded and regulated by governments (such as legal gender, monetary income or profession).

In contrast, characteristics like sexual orientation and gender identity are frequently \textit{unobserved} \citep{andrus2020we, crocker1998social, jacobs2019measurement}.
Multiple factors contribute to this lack of data. In some cases, the plan for data collection fails to incorporate questions on sexual orientation and gender identity---potentially because the data collector did not consider or realise that they are important attributes to record~\citep{herek1991avoiding}. As a result, researchers may inherit datasets where assessment of sexual orientation and gender identity is \textit{logistically excluded}.
In other situations, regardless of the surveyor's intent, the collection of certain personal data may threaten an individual's privacy or their safety.
Many countries have legislation that actively discriminates against LGBTQ+ people~\citep{humanrightswatchmaps}.
Even in nations with hard-won protections for the queer community, cultural bias persists. To shield individuals from this bias and protect their privacy, governments may instate legal protections for sensitive data, including sexual orientation~\citep{europeancommissionwhat}.
As a result, such data may be \textit{ethically or legally precluded} for researchers.
Finally, as recognised by discursive theories of gender and sexuality, sexual orientation and gender identity are fluid cultural constructs that may change over time and across social contexts~\citep{butler2011bodies}. Attempts to categorise, label, and record such information may be inherently ill-posed \citep{hamidi2018gender}. Thus, some characteristics are unobserved because they are \textit{fundamentally unmeasurable}. These inconsistencies in awareness and measurability yield discrepancies and tension in how fairness is applied across different contexts \citep{bogen2020awareness}.

Studies of race and ethnicity are not immune to these challenges. Race and ethnicity may be subject to legal observability issues in settings where race-based discrimination is a sensitive issue (e.g., hiring). Additionally, the definition of racial and ethnic groups has fluctuated across time and place \citep{hanna2020towards}. This is exemplified by the construction of Hispanic identity in the United States and its inclusion on the National Census, as well as the exclusion of multiracial individuals from many censuses until relatively recently~\citep{mora2014making}.
Though we choose to focus our analysis on queer identity, we note that the observability and measurability of race are also important topics (e.g., \citep{scheuerman2020we}).

\section{Areas for Future Research}

The field of algorithmic fairness in machine learning is rapidly expanding. To date, however, most studies have overlooked the implications of their work for queer people. To include sexual orientation and gender identity in fairness research, it will be necessary to explore new technical approaches and evaluative frameworks. To prevent the risk of AI systems harming the queer community---as well as other marginalised groups whose defining features are similarly unobserved and unmeasurable---fairness research must be expanded.

\subsection{Expanding Fairness for Queer Identities}

Machine learning models cannot be considered fair unless they explicitly factor in and account for fairness towards the LGBTQ+ community.
To minimise the risks and harms to queer people worldwide and avoid contributing to ongoing erasures of queer identity, researchers must propose solutions that explicitly account for fairness with respect to the queer community. %

The intersectional nature of sexual orientation and gender identity~\citep{parent2013approaches} emerges as a recurring theme in our discussions of online abuse, health and employment. These identities cannot be understood without incorporating notions of economic and racial justice. Deployed AI systems may pose divergent risks to different queer subcommunities; AI risks may vary between gay, bisexual, lesbian, transgender and other groups. It is therefore important to apply an appropriate level of granularity to the analysis of fairness for algorithmic issues.
Policies can simultaneously improve the position of certain queer groups while adversely affecting others---highlighting the need for an intersectional analysis of queer fairness.

Demographic parity has been the focus of numerous ML fairness studies and seems to closely match people's conceptions of fairness~\citep{srivastava2019mathematical}. However, this idea is hard to promote in the context of queer ML fairness. Substantial challenges are posed by the sensitivity of group membership information and its absence from most research datasets, as well as the associated outing risks associated with attempts to automatically derive such information from existing data~\citep{bjarnadottir2019nurse, ehrenfeld2019development, young2017toward}. Consensually provided self-identification data, if and when available, may only capture a fraction of the community. The resulting biased estimates of queer fairness may involve high levels of uncertainty~\citep{ethayarajh2020your}, though it may be possible to utilise unlabeled data for tightening the bounds~\citep{ji2020can}. While it is possible to root the analysis in proxy groups~\citep{gupta2018proxy}, there is a risk of incorporating harmful stereotypes in proxy group definitions, potentially resulting in harms of representation~\citep{abbasi2019fairness}. Consequently, most ML fairness solutions developed with a specific notion of demographic parity in mind may be inappropriate for ensuring queer ML fairness.

Individual~\citep{dwork2012fairness, jung2019eliciting}, counterfactual~\citep{kusner2017counterfactual}, and contrastive~\citep{chakraborti2020contrastive} fairness present alternative definitions and measurement frameworks that may prove useful for improving ML fairness for queer communities. However, more research is needed to overcome implementational challenges for these frameworks and facilitate their adoption.

A small body of work aims to address fairness for protected groups when the collection of protected attributes is legally precluded (e.g., by privacy and other regulation). Adversarially Re-weighted Learning~\citep{lahoti2020fairness} aims to address this issue by relying on measurable covariates of protected characteristics (e.g., zip code as a proxy for race). This approach aims to achieve intersectional fairness by optimising group fairness between all computationally identifiable groups~\citep{kearns2018preventing, kim2018fairness}. Distributionally robust optimisation represents an alternative method for preventing disparity amplification, bounding the worst-case risk over groups with unknown group membership by optimising the worst-case risk over an appropriate risk region~\citep{hashimoto2018fairness}. These methods have helped establish a link between robustness and fairness, and have drawn attention to the synergistic benefits of considering the relationship between fairness and ML generalisation~\cite{creager2020exchanging}. Other adversarial approaches have also been proposed for improving counterfactual fairness, and by operating in continuous settings, have been shown to be a better fit for protected characteristics that are hard to enumerate~\citep{grari2020adversarial}.

Fairness mitigation methods have been shown to be vulnerable to membership inference attacks where the information leak increases disproportionately for underprivileged subgroups~\citep{chang2020privacy}. This further highlights the tension between privacy and fairness, a common theme when considering the impact of AI systems of queer communities. It is important to recognise the need for fairness solutions to respect and maintain the privacy of queer individuals and to be implemented in a way that minimises the associated reidentifiability risks. Differentially private fair machine learning~\citep{jagielski2019differentially} could potentially provide such guarantees, simultaneously meeting the requirements of fairness, privacy and accuracy.

Putting a greater emphasis on model explainability may prove crucial for ensuring ethical and fair AI applications in cases when fairness metrics are hard or impossible to reliably compute for queer communities. Understanding how AI systems operate may help identify harmful biases that are likely to have adverse downstream consequences, even if these consequences are hard to quantify accurately. Even in cases when queer fairness can be explicitly measured, there is value in identifying which input features contribute the most to unfair model outcomes~\citep{begley2020explainability}, in order to better inform mitigation strategies.

It is important to acknowledge the unquestionable cisnormativity of sex and gender categories traditionally used in the AI research literature. The assumption of fixed, binary genders fails to include and properly account for non-binary identities and trans people~\citep{keyes2018misgendering}. Incorporating such biases in the early stages of AI system design poses a substantial risk of harm to queer people. Moving forward, more attention should be directed to address this lacuna.

Creating more-equitable AI systems will prove impossible without listening to those who are at greatest risk.
Therefore, it is crucial for the AI community to involve more queer voices in the development of AI systems, ML fairness, and ethics research~\citep{arnstein1969ladder, poulsen2020queering}. 
For example, the inclusion of queer perspectives might have prevented the development of natural language systems that inadvertently censor content which is wrongly flagged as abusive or inappropriate simply due to its adjacency to queer culture, such as in the example of scoring drag queen language as toxic.
Researchers should make efforts to provide a safe space for LGBTQ+ individuals to express their opinions and share their experiences.
Queer in AI workshops have recently been organised at the Neural Information Processing Systems conference~\citep{queerinaineurips2018, queerinaineurips2019} and the International Conference on Machine Learning~\citep{queerinaiicml2020}, providing a valuable opportunity for queer AI researchers to network in a safe environment and discuss research at the intersection of AI and queer identity.

\subsection{Fairness for Other Unobserved Characteristics}

The queer community is not the only marginalised group for which group membership may be unobserved \citep{crocker1998social}. Religion, disability status, and class are additional examples where fairness is often challenged by observability \citep{kattari2018you, sanchez2001passing}. Critically, they may also benefit from developments or solutions within queer fairness research. For example, in nations where individuals of certain religious groups are persecuted or subjected to surveillance, privacy is an essential prerequisite for safety. Persecution targeting religious communities may also include censorship or manipulation of information~\citep{cook2017battle}.
Even in nations where religious freedoms are legally protected, religious minorities may be subjected to online abuse such as hate speech or fear-mongering stereotypes~\citep{awan2014islamophobia}.

Although the nature of the discrimination is different, people with disabilities are also a frequent target of derogatory language on the internet, and are more likely to be harassed, stalked or trolled online, often to the detriment of their mental health~\citep{sherry2019disablist}.
Youth with disabilities more frequently suffer from adverse mental health due to bullying, and people of all ages with physical disabilities are at higher risk for depression~\citep{king2018extent, turner1988physical}.
Therefore, individuals with disabilities may benefit from insights on the interaction of unobserved characteristics and mental health.
Lower-income and lower-class individuals also suffer from worse mental health, particularly in countries with high economic inequality~\citep{liu2008social}.
Fairness for class and socioeconomic status is also an important consideration for employment, where class bias in hiring limits employee diversity and may prevent economic mobility~\citep{kraus2019evidence}.

Any particular dataset or AI application may instantiate observability difficulties with respect to multiple demographics. This may frequently be the case for disability status and class, for example. Individual fairness---a set of approaches based on the notion of treating similar individuals similarly~\citep{dwork2012fairness, jung2019eliciting}---could potentially promote fairness across multiple demographics. These approaches entail a handful of challenges, however. The unobserved group memberships cannot be incorporated in the similarity measure. As a result, the similarity measure used for assessing individual fairness must be designed carefully. To optimise fairness across multiple demographics and better capture the similarity between people on a fine-grained level, similarity measures will likely need to incorporate a large number of proxy features. This would be a marked divergence from the proposed measures in most published work. Counterfactual and contrastive fairness metrics come with their own set of practical implementation challenges.

On the other hand, the approaches aimed at providing worst-case fairness guarantees for groups with unknown %
group membership~\citep{hashimoto2018fairness, lahoti2020fairness} apply by definition to any marginalised group. They are also specifically tailored to address the situation of unobserved protected characteristics. Therefore, fairness solutions required to address queer ML fairness are likely to be applicable to other groups as well.

Fairness challenges are institutionally and contextually grounded, and it is important to go beyond purely computational approaches to fully assess the sociotechnical aspects of the technology being deployed. The complexity of these issues preclude any single group from tackling them in their entirety, and a resolution would ultimately require an ecosystem involving a multitude of partnering organisations, jointly monitoring, measuring and reporting fairness of such systems~\citep{veale2017fairer}.

These issues are only a small sample of the common challenges faced by groups with typically unobserved characteristics.
We invite future work to explore the impact of AI from the perspective of such groups. It is important to acknowledge that people with different identities have distinct experiences of marginalisation, stigmatisation and discrimination. However, recognising common patterns of injustice will likely enable the development of techniques that can transfer across communities and enhance fairness for multiple groups.
In this way, shared ethical and technical design principles for AI fairness will hopefully result in a more equitable future.

\section{Conclusion}

The queer community has surmounted numerous historical challenges and continues to resist oppression in physical and digital spaces around the world. Advances in artificial intelligence represent both a potential aid to this resistance and a risk of exacerbating existing inequalities. This risk should motivate researchers to design and develop AI systems with fairness for queer identities in mind. Systems that attempt to label sexual orientation and gender identity, even for the purpose of fairness, raise technical and ethical challenges regarding observability and measurability.

A new discourse on queer fairness has the potential to identify moral and practical considerations shared across queer communities, as well as concerns specific to particular subpopulations in particular places. By further developing techniques supporting fairness for unobserved characteristics, the machine learning community can support queer communities and other marginalised groups. Broadly, the present work---surveying the ways in which AI may ameliorate or exacerbate issues faced by queer communities%
---emphasises the need for machine learning practitioners to design systems with fairness and dignity in mind.

\begin{acks}
We would like to thank Aliya Ahmed, Dorothy Chou, Ben Coppin, Michelle Dunlop, Thore Graepel, William Isaac, Koray Kavukcuoglu, Guy Scully, and Laura Weidinger for the support and insightful feedback that they provided for this paper.

Any opinions presented in this paper represent the personal views of the authors and do not necessarily reflect the official policies or positions of their organisations.
\end{acks}

\bibliographystyle{ACM-Reference-Format}
\bibliography{main}

\end{document}